\documentclass[twocolumn,pra,psfig,showpacs,superscriptaddress]{revtex4}

\usepackage{graphicx}
\usepackage{times}

\usepackage[all]{xy}
\usepackage{amsmath}

\usepackage{graphics}

\usepackage{amssymb}

\begin{document}

\title{Phase transition in the massive Gross-Neveu model in toroidal
topologies}

\author{F. C. Khanna{\footnote {fkhanna@ualberta.ca}}}
\affiliation{Theoretical Physics Institute, University of Alberta,
Edmonton, Alberta T6G 2J1, Canada} \affiliation{TRIUMF, 4004,
Westbrook Mall, Vancouver, British Columbia V6T 2A3, Canada}
\author{A. P. C. Malbouisson{\footnote {adolfo@cbpf.br}}}
\affiliation{Centro Brasileiro de Pesquisas F\'{\i}sicas/MCT,
22290-180, Rio de Janeiro, RJ, Brazil}
\author{J. M. C. Malbouisson{\footnote {jmalboui@ufba.br}}}
\affiliation{Instituto de F\'{\i}sica, Universidade Federal da
Bahia, 40210-340, Salvador, BA, Brazil} \affiliation{Departamento de
F\'{\i}sica, Faculdade de Ci\^encias, Universidade do Porto,
4169-007, Porto, Portugal}
\author{A. E. Santana{\footnote {asantana@fis.unb.br}}}
\affiliation{Instituto de F\'{\i}sica, International Center of
Physics, Universidade de Bras\'{\i}lia, 70910-900, Bras\'\i lia, DF,
Brazil}

\begin{abstract}
We use methods of quantum field theory in toroidal topologies to
study the $N$-component $D$-dimensional massive Gross-Neveu model,
at zero and finite temperature, with compactified spatial
coordinates. We discuss the behavior of the large-$N$ coupling
constant ($g$), investigating its dependence on the compactification
length ($L$) and the temperature ($T$). For all values of the fixed
coupling constant ($\lambda$), we find an asymptotic-freedom type of
behavior, with $g\rightarrow 0$ as $L\rightarrow 0$ and/or
$T\rightarrow \infty$. At $T=0$, and for $\lambda \geq
\lambda_{c}^{(D)}$ (the strong coupling regime), we show that,
starting in the region of asymptotic freedom and increasing $L$, a
divergence of $g$ appears at a finite value of $L$, signaling the
existence of a phase transition with the system getting spatially
confined. Such a spatial confinement is destroyed by raising the
temperature. The confining length, $L_{c}^{(D)}$, and the
deconfining temperature, $T_{d}^{(D)}$, are determined as functions
of $\lambda$ and the mass ($m$) of the fermions, in the case of
$D=2,3,4$. Taking $m$ as the constituent quark mass ($\approx 350\:
MeV$), the results obtained are of the same order of magnitude as
the diameter ($\approx 1.7\, fm$) and the estimated deconfining
temperature ($\approx 200\: MeV$) of hadrons.
\end{abstract}

\pacs{11.10.Kk; 11.10.Wx}

\maketitle

\section{Introduction}

The strong interaction among quarks and gluons, the constituents of
the hadronic matter, has such a structure that obligates them to
live spatially confined, at low temperatures, within distances $\sim
1\,{\rm fm}$ in colorless states. It is usually accepted that in an
earlier stage of the Universe, as it cooled down, quarks and gluons
condensed into hadrons at an estimated temperature of the order of
$200\,{\rm MeV}$. At very high energies, deep inelastic scattering
indicates that the quarks are nearly free, a regime denominated as
asymptotic freedom.

In the standard model, the theory of strong interactions is quantum
chomodynamics (QCD), which should describe such facts, accounting
also for the nuclear forces. However, QCD has a very involved
mathematical structure, which practically prevents us from finding
analytical results taking into account both confinement and
asymptotic freedom. Lattice calculations have been implemented to
simulate the behavior of the theory in the confining region, both at
zero and finite temperature, providing (among other results) an
estimate of the deconfining temperature. Rigorous QCD calculations,
both at zero and finite temperature, have been worked
out~\cite{GPY,Kalash} but mainly treating the asymptotically free
domain at high energies or high temperatures, where perturbation
theory is applicable.

Due to the difficulty of treating QCD analytically, phenomenological
approaches and studies of effective, simplified, models have been
stimulated along the years to give clues to the behavior of hadronic
systems. A celebrated effective model, which shares with QCD some
basic properties, is the Nambu-Jona-Lasinio (NJL) model~\cite{NJL}.
One of its sectors, providing the simplest effective model which may
be considered as describing quark interactions, is a direct
four-fermion coupling, where gluon fields and color degrees of
freedom are integrated out, resembling the Fermi treatment of the
weak interaction. This corresponds to the Gross-Neveu (GN)
model~\cite{GN}, considered in space-time dimension $D=4$. Although
the GN model is not renormalizable for dimensions greater than
$D=2$, the Euclidian model has been shown to exist and has been
constructed for $D=3$ in the large-$N$ limit \cite{Jarrao}. But,
within the spirit of effective theories, perturbative
renormalizability is not a requirement to have a physically
meaningful model~\cite{parisi,anp,Gaw1,rwp1,Weinberg1}.

The GN model, as a prototype model for interacting fermions, has
been analyzed extensively in recent
years~\cite{gn111,gn2,gn3,gn5,gn9,gn10,gn12,gn13,gn14,gn19},
including the study  of continuous and discrete chiral
symmetry~\cite{gn7,gn8}. For the version with $N$ massless fermions
in (2+1)-dimensions, for instance, a chiral symmetry breaking is
found in perturbative analysis, with the restoration of such a
symmetry at finite temperature~\cite{gn15}. In particle physics,
these results provide insights into the intricate structure of the
hadronic matter, such as for the quark confinement/deconfinement
phase transition~\cite{gn6,gn18}.

The four-point contact interaction of the GN model is similar to the
delta interaction in the BCS theory of superconductivity. In the
latter case, as in other systems of condensed matter, the
susceptibility arising from the linear response theory has a
divergence at a finite temperature indicating the existence of a
second-order phase transition between a disordered and a condensed
phase~\cite{Doniach}. Spontaneous symmetry breaking is the common
feature underlying all these phenomena~\cite{Weinberg2}.
Recently~\cite{KMMSepl}, we have shown that such instability appears
in the one-component massive tridimensional GN model at finite
temperature. Here, we intend to use similar treatment to investigate
the existence of a phase transition in the massive, $N$-component,
GN model.

We employ methods of quantum field theory in toroidal
topologies~\cite{BF,Polch,AW,AP09,Book,AP11} to extend previous
results~\cite{GNN,GNNT} for the Euclidian massive, $N$-component, GN
model in the large-$N$ limit. This amounts to consider the GN model
in a $D$-dimensional space-time and to compactify a $d$-dimensional
($d\leq D$) subspace. The compactification is a generalization of
the Matsubara procedure. The Matsubara imaginary-time formalism
corresponds to considering fields in a space with topology
$\mathbb{S}^{1}\times \mathbb{R}^{D-1}$, where $\mathbb{S}^{1}$ is a
circumference of length $\beta$, with periodic (anti-periodic)
boundary conditions for bosons (fermions). Such a compactification
of the Euclidian time can be directly generalized to include the
compactification of space coordinates as well. This allows us to
consider field theoretical models with spatial constraints, at zero
or finite temperature, by using generating functionals with a
path-integral formalism on the topology $\mathbb{S}^{1_1}\times
\cdots \times \mathbb{S}^{1_d}\times
\mathbb{R}^{D-d}$~\cite{BF,Polch,AW,AP09}. These ideas have been
established recently on a firm foundation~\cite{Book,AP11} and
applied in different physical situations, for example: for
spontaneous symmetry breaking in the compactified $\phi^4$
model~\cite{Ra,boson,NPB}; for second-order phase transitions in
superconducting films, wires and grains~\cite{MMSK,JMP,JMP2}; for
the Casimir effect for bosons and
fermions~\cite{Ford,KZ,kha1,Cef,AN-Z,TC}; for size effects in the
NJL model~\cite{KH,LMA,EKTZ,LMA1,LMA2}; and, for electrodynamics
with an extra dimension~\cite{FAR}.

We treat particularly the cases $D=2,3,4$ with all spatial
dimensions compactified, initially at zero temperature and then
discuss temperature effects by compactifying the imaginary time in a
length $\beta = T^{-1}$, $T$ being the temperature. This corresponds
to considering the system contained in a parallelepiped box, with
anti-periodic boundary conditions on its faces, at finite $T$. We
study the behavior of the system as a function of its size and of
the temperature, concentrating on the dependence of the large-$N$
coupling constant on the compactification length and temperature. We
show that, even at $T=0$, a singularity in the $4$-point function
may appear driven by changes in the compactification length
suggesting the existence of a second-order phase transition in the
system. This can be interpreted as a spatial confinement transition,
which may be present in the massive version of the GN model.

We use concurrently dimensional and analytic regularizations and
employ a subtraction scheme where the polar terms, arising from
Epstein--Hurwitz generalized $zeta$-functions, are suppressed.
Results obtained with this procedure have similar structure for all
values of $D$, which gives us confidence that they are meaningful
for the $4$-dimensional space-time. This is reinforced {\it a
posteriori} by the fact that the numerical results found for $D=4$
are of the same order of magnitude as the corresponding values for
$D=2$ and $D=3$. For the massive GN model in the large-$N$ limit,
discussed in the present paper, we obtain simultaneously asymptotic
freedom type of behavior and spatial confinement, in the strong
coupling regime, for low temperatures. We also show that, as the
temperature is increased, a deconfining transition occurs. We
calculate the values of the confining lengths and the deconfining
temperatures for $D=2,3,4$.

This article starts by discussing, in Sec.~2, the $D$-dimensional
massive, $N$-component, Gross-Neveu model with $d$ ($\leq D$)
compactified dimensions. In Sec.~3, the calculation of the effective
large-$N$ coupling constant is carried out for the cases with
$D=2,3,4$ at zero temperature. Temperature effects are presented in
Sec.~4. The last section provides a comparison of our estimated
confining lengths and deconfining temperatures with experimental
values.

\section{Compactified Gross-Neveu model}

The massive GN model in a $D$-dimensional Euclidean space is
described by the Wick-ordered Lagrangian density
\begin{equation}
\mathcal{L}=:\bar{\psi}(x)(i{\not{\hbox{\kern-2.3pt $\nabla$}}}
+m)\psi (x):+\frac{u}{2}(:\bar{\psi} (x)\psi (x):)^{2},  \label{GN}
\end{equation}
where $m$ is the mass, $u$ is the coupling constant, $x$ is a point
of $\mathbb{ R}^{D}$ and the $\gamma $'s are the Dirac matrices. We
consider the GN model in its $N$-component version, so that $\psi
(x)$ represents a spin $\frac{1}{2}$ field having $N$ (flavor)
components, $\psi ^{a}(x)$, $a=1,2,...,N$, with summations over
flavor and spin indices being understood in Eq.~(\ref{GN}). We
calculate quantities of interest by taking the large-$N$ limit where
$N\rightarrow\infty$ and $u \rightarrow 0$ in such way that $N u =
\lambda$ remains finite. Throughout the text we use natural units
with $\hbar=\,c=\,k_B=\,1$.

Our main goal is to determine the large-$N$ ({\it effective})
coupling constant when $d$ ($\leq D$) Euclidian coordinates, say
$x_1,\dots,x_d$, are compactified, that is, considering the system
in a topology $\mathbb{S}^{1_1}\times \cdots \mathbb{S}^{1_d}\times
\mathbb{R}^{D-d}$. This corresponds to restricting the coordinates
$x_i$ to segments of length $L_i$ ($i=1,2,....d$), with the
fermionic field $\psi(x)$ satisfying anti-periodic boundary
conditions. If all $x_i$ are spatial coordinates, the model refers
to the system compactified in a $d$-dimensional box at zero
temperature while, with one coordinate being the Euclidian time (say
$x_d$), one has the system with $d-1$ compactified spatial
dimensions at finite temperature; in this latter case, $L_d$ would
stand for $\beta=1/T$, the inverse of the temperature. For massless
fermions, this spatial compactification with anti-periodic boundary
conditions is equivalent to considering the system constrained to
``live" inside a parallelepiped ``box", with edges $L_i$
($i=1,2,....d$), under bag model conditions (no outgoing currents)
on parallel, opposite, faces \cite{bag1,bag2}. In our case, to
calculate $n$-point functions, we apply the generalized Matsubara
prescription, which amounts to modifying the Feynman rules
performing the replacements
\begin{subequations}\label{Matsufreq}
\begin{equation}
 k_{i} \; \rightarrow \;  \nu_{i} =
 \frac{2\pi (n_i+\frac{1}{2})}{L_i}\; ,\;\;
 i=1,2,...,d \; ;
\end{equation}
\begin{eqnarray}
\int \frac{d ^D {k}}{(2\pi)^D }\, F ( {k} ) \, &
 \rightarrow \, &
\frac{1}{L_{1}\ldots L_d} \nonumber \\
 & & \times \sum _{ \{n_{i} \} = - \infty
}^{+\infty} \int \frac{d ^{D - d }{\bf k}}{(2\pi)^{D-d} }\, F (
\{k_{i}\},{\bf k} )\, , \nonumber \\
 & &
\end{eqnarray}
\end{subequations}
where $ \{n _i\} = \{n_1 , \ldots, n _d\}$, with $n_i \in {\mathbb
Z}$, and ${\bf k}$ is a $(D-d)$-dimensional vector in momentum
space; the discreet momenta  $\nu_{i}$ are referred to as Matsubara
frequencies. Additionally, it should be pointed out that the choice
of anti-periodic boundary conditions for the spatial
compactification, instead of the simpler periodic ones, is also due
to the fact that they emerge naturally in the generalization of the
KMS (Kubo-Martin-Schwinger) conditions satisfied by correlation
functions for fermionic fields in toroidal
topologies~\cite{Book,AP11}.

We shall define the large-$N$ effective coupling constant between
the fermions in terms of the $4$-point function at zero external
momenta. The $\{ L_i\}$-dependent four-point function, at leading
order in $\frac{1}{N}$, is given by the sum of chains of one-loop
(bubble) diagrams, which can be formally expressed as
\begin{equation}
\Gamma _{Dd}^{(4)}(0;\{L_i\},u)=\;\frac{u}{1+Nu\Pi_{Dd}(\{L_i\})} ,
\label{4-point1}
\end{equation}
where the $\{L_i\}$-dependent one-loop Feynman diagram is given by
\begin{widetext}
\begin{equation}
\Pi_{Dd}(\{L_i\}) = \frac{1}{L_1\cdots L_d} \sum_{\{n_i\}=-\infty
}^{\infty} \int \frac{d^{D-d} {\bf k}}{(2\pi)^{D-d}} \left[
\frac{m^{2}-{\bf k}^2-\sum_{i=1}^{d}{\nu}_{i}^{2}} {\left({\bf k}^2
+\sum_{i=1}^{d}{\nu}_{i}^{2}+m^{2}\right)^{2}}\right] .
\label{sigma0}
\end{equation}
\end{widetext}

Prior to defining an effective large-$N$ coupling constant, we have
to deal with the ultraviolet divergences of $\Pi_{Dd}(\{L_i\})$. To
simplify the use of regularization techniques, we introduce the
dimensionless quantities $b_i=(m L_i )^{-2}$ ($i=1,\dots ,d$) and
$q_{j}=k_{j}/2\pi m$ $(j=d+1,\dots ,D)$, in terms of which the
one-loop diagram is written as
\begin{widetext}
\begin{equation}
\Pi_{Dd}(\{b_i\})  =  \left. \Pi_{Dd}(s;\{b_i\}) \right|_{s=2} =
\frac{m^{D-2}}{4\pi^2}\sqrt{b_1\cdots
b_d}\left.\left\{\frac{1}{2\pi^2} U_{Dd}(s;\{b_i\}) -
U_{Dd}(s-1;\{b_i\}) \right\}\right|_{s=2} , \label{sigma2}
\end{equation}
where
\begin{equation}
U_{Dd}(\mu;\{b_i\}) = \sum_{\{n_i\}=-\infty }^{\infty} \int
\frac{d^{D-d} {\bf q}}{\left[{\bf q}^2
+\sum_{j=1}^{d}b_j(n_{j}+\frac{1}{2})^{2}+(2\pi)^{-2}\right]^{\mu}}
. \label{UDd}
\end{equation}
\end{widetext}
We find from Eq.~(\ref{sigma2}) that $\Pi_{Dd}$ has dimension of
$m^{D-2}$, which is inverse of the mass dimension of the coupling
constant.

We employ a modified minimal subtraction scheme which uses
concurrently dimensional and analytical regularizations. In this
scheme, the subtracted terms are poles (for even $D\geq 2$) of the
Epstein-Hurwitz $zeta-$functions. First, using well-known
dimensional regularization formulas to perform the integral over
${\bf q}=(q_{d+1},\dots,q_D)$ in Eq.~(\ref{UDd}), we obtain
\begin{eqnarray}
U_{Dd}(\mu;\{b_i\}) & = & \pi^{\frac{D-d}{2}}\,\frac{\Gamma(\mu -
\frac{D-d}{2})}{\Gamma(\mu)} \sum_{\{n_i\}=-\infty
}^{\infty} \nonumber \\
 & & \times  \left[\sum_{j=1}^{d}b_j\left(n_{j}+\frac{1}{2}\right)^{2}
+(2\pi)^{-2}\right]^{\frac{D-d}{2}-\mu} .  \nonumber \\
 & & \label{UDd2}
\end{eqnarray}
The summations over half-integers in this expression can be
transformed into sums over integers leading to
\begin{eqnarray}
U_{Dd}(\mu;\{b_i\}) & = & \pi^{\frac{D-d}{2}} \, \frac{\Gamma
(\mu-\frac{(D-d)}{ 2})}{\Gamma (\mu)}\, 4^{\eta } \nonumber \\
 & & \times \left[ Z_{d}^{h^{2}}(\eta ,b_1,\dots,b_d)  \right. \nonumber \\
 & &
 - \sum_{i=1}^{d}Z_{d}^{h^{2}}(\eta ,\dots,4b_i,\dots) \nonumber \\
 & &  + \sum_{i<j=1}^{d} Z_{d}^{h^{2}}(\eta ,\dots, 4b_i,\dots,4b_j,
\dots)  \nonumber \\
 & & - \cdots + \left.  (-1)^{d}\, Z_{d}^{h^{2}}(\eta
,4b_1,\dots,4b_d)\right] , \nonumber \\
 & & \label{UDd3}
\end{eqnarray}
where $h^2=\pi^{-2}$, $\,\eta =\mu-\frac{D-d}{2}\,$ and
\begin{equation}
Z_{d}^{h^{2}}(\eta,\{a_i\})=\sum_{\{n_i\}=-\infty }^{\infty }
\left[\sum_{j=1}^{d} a_j n_{j}^{2}+h^{2}\right]^{-\eta}
\end{equation}
is the multiple ($d$-dimensional) Epstein-Hurwitz $zeta$-function.

The Epstein-Hurwitz $zeta$-function $Z_{d}^{h^{2}}(\eta,\{a_i\})$
can be analytically extended to the whole complex $\eta$-plane
\cite{NPB}, through a generalization of the procedure presented in
Refs.~\cite{ER,K}; then we find
\begin{widetext}
\begin{eqnarray}
Z_{d}^{h^{2}}(\eta,\{a_i\})
&=&\frac{\pi^{\frac{d}{2}}}{\sqrt{a_1\cdots a_d}\,\Gamma (\eta)}
\left[ \frac{1}{h^{2(\eta-d)}} \Gamma \left(\eta-\frac{d}{2}\right)
\right.  \nonumber \\
 & & +\, \sum_{\theta=1}^{d} 2^{\theta+1}
\sum_{\{\sigma_{\theta}\}}
 \sum_{\{n_{\sigma_{\theta}}\}=1}^{\infty}\left( \frac{\pi
}{h}\sqrt{\frac{n_{\sigma_{1}}^{2}}{a_{\sigma_1}} + \cdots +
\frac{n_{\sigma_{\theta}}^{2}}{a_{\sigma_{\theta}}}}\right) ^{\eta
-\frac{d}{2}} \left. K_{\eta -\frac{d}{2}}\left( 2\pi
h\sqrt{\frac{n_{\sigma_{1}}^{2}}{a_{\sigma_1}} + \cdots
+\frac{n_{\sigma_{\theta}}^{2}}{a_{\sigma_{\theta}}}}\right) \right]
, \label{Z2}
\end{eqnarray}
\end{widetext}
where $\{\sigma_{\theta}\}$ represents the set of all combinations
of the indices $\{1,2,\dots,d\}$ with $\theta$ elements and
$K_{\alpha }(z)$ is the Bessel function of the third kind.
Consequently, the function $U_{Dd}(\mu;\{b_i\})$ can also be
analytically continued to the whole complex $\mu$-plane.

Taking $Z_{d}^{h^{2}}(\eta,\{a_i\})$ given by Eq.~(\ref{Z2}),
grouping similar terms appearing in the parcels of Eq.~(\ref{UDd3})
and using the identity
\begin{equation}
\sum_{j=1}^{N}\left( \frac{-1}{2} \right)^j \frac{N!}{j!(N-j)!} =
\frac{1}{2^N}\, ,
\end{equation}
\begin{widetext}
\noindent we obtain
\begin{equation}
U_{Dd}(\mu;\{b_i\})  =  \frac{2^{2\mu-D}\pi^{2\mu-\frac{D}{2}}}
{\Gamma(\mu)} \frac{1}{\sqrt{b_1\cdots b_d}} \left[ \Gamma\left( \mu
- \frac{D}{2} \right) + 2^{\frac{D}{2}}\, W_{Dd}(\mu;\{ b_i \})
\right]  \label{UDd4}
\end{equation}
with $W_{Dd}(\mu;\{ b_i \})$ given by
\begin{equation}
W_{Dd}(\mu;\{ b_i \})  =  2^{1 - \mu} \sum_{j=1}^{d} 2^{2 j}
 \sum_{\{ \rho_j \}} \sum_{\{ c_{\rho_k}=1,4 \}} \left(
\prod_{k=1}^{j} \frac{(-1)^{c_{\rho_k} - 1}}{\sqrt{c_{\rho_k}}}
\right) \,
F_{Dj}(\mu;c_{\rho_1}b_{\rho_1},\dots,c_{\rho_j}b_{\rho_j})  ,
\label{WDd}
\end{equation}
where $\{ \rho_j \}$ stands for the set of all combinations of the
indices $\{1,2,\dots,d\}$ with $j$ elements and the functions
$F_{Dj}(\mu;a_1,\dots,a_j)$, for $j=1,\dots,d$, are defined by
\begin{equation}
F_{Dj}(\mu;a_1,\dots,a_j)  =  \sum_{n_1,\dots,n_j=1}^{\infty}\left(
2 \sqrt{\frac{n_{1}^{2}}{a_{1}} + \cdots +
\frac{n_{j}^{2}}{a_{j}}}\right) ^{\mu - \frac{D}{2}} K_{\mu -
\frac{D}{2}}\left( 2 \sqrt{\frac{n_{1}^{2}}{a_{1}} + \cdots
+\frac{n_{j}^{2}}{a_{j}}}\right)  . \label{FDj}
\end{equation}
\end{widetext}

Substituting Eq.~(\ref{UDd4}) into Eq.~(\ref{sigma2}) leads directly
to an analytic extension of $\Pi_{Dd}(s;\{b_i\})$ for complex values
of $s$, in the vicinity of $s=2$. In fact, $\Pi_{Dd}(s;\{b_i\})$ can
be written as
\begin{eqnarray}
\Pi_{Dd}(s;\{b_i\}) & = & \Pi_{Dd}^{{\rm polar}}(s)
 +  \frac{m^{D-2}}{(2\pi)^{\frac{D}{2}-2s+4}\,\Gamma(s)}  \nonumber \\
  & & \times \left[ 2\, W_{Dd}(s;\{ b_i \}) \right. \nonumber \\
  & & - \, \left. (s-1)\,
 W_{Dd}(s-1;\{ b_i \}) \right]  , \label{SigmaS}
\end{eqnarray}
where
\begin{equation}
\Pi_{Dd}^{{\rm polar}}(s) =
\frac{m^{D-2}\,\pi^{\frac{D}{2}}(s-1-D)}{(2\pi)^{D-2s+4}\,\Gamma(s)}
 \, \Gamma\left( s - 1 - \frac{D}{2} \right)
 \label{SigmaPolar}
\end{equation}
and the functions $W_{Dd}(\mu;\{ b_i \})$ are given by
Eq.~(\ref{WDd}). We notice that the first term in this expression
for $\Pi_{Dd}(s;\{b_i\})$, $\Pi_{Dd}^{{\rm polar}}(s)$, does not
depend on parameters $b_i$, that is, it is independent of the
compactification lengths $L_i$ ($i=1,\dots,d$). At $s=2$, because of
the poles of the $\Gamma$-function, such a term is divergent for
even dimensions $D\geq 2$.

In order to obtain a finite single bubble function, we shall use a
modified minimal subtraction scheme, where terms to be subtracted
have poles appearing at the physical value $s=2$. Thus, the polar
parcel given by Eq.~(\ref{SigmaPolar}) will be suppressed and, for
the sake of uniformity, this term is also subtracted in the case of
odd dimensions, where no pole of the $\Gamma$-function is present;
in such a situation, we perform a finite subtraction. In this way,
using the same notation, we define the finite one-loop diagram by
the relation
\begin{equation}
\Pi_{Dd}(\{b_i\}) = \left. \left\{ \Pi_{Dd}(s;\{b_i\}) -
\Pi_{Dd}^{\rm polar}(s) \right\} \right|_{s=2} . \label{RenSigma}
\end{equation}
Therefore, the finite one-loop diagram, which depends on the
compactification lengths $L_i$ and arises from the regular part of
the analytical extension of the Epstein-Hurwitz $zeta$-functions, is
given by
\begin{equation}
\Pi_{Dd}(\{b_i\}) =  \frac{m^{D-2}}{(2\pi)^{\frac{D}{2}}}\, \left[
2\, W_{Dd}(2;\{ b_i \}) - W_{Dd}(1;\{ b_i \}) \right]  .
\label{SigmaR}
\end{equation}
From now on, we shall deal only with finite quantities that are
obtained following this subtraction prescription. Notice that,
replacing $b_i$ by $(m L_i)^{-2}$ in the above expression, we
recover explicitly $\Pi_{Dd}(\{L_i\})$. Now, we proceed to analyze
the behavior of the large-$N$ coupling constant in various cases.

\section{Coupling constant in the large-$N$ limit}

In field theories with four-fermion interactions, the coupling
constant is defined in terms of the four-point function at fixed
external momenta; here we choose ${\bf p} = 0$. In this situation,
the coupling constant can be interpreted as measuring the strength
of the interaction between the fermions. The large-$N$
($\{b_i\}$-dependent) coupling constant, for $d$ ($\leq D$)
compactified dimensions, is then obtained by substituting
$\Pi_{Dd}(\{b_i\})$ into Eq.~(\ref{4-point1}) and taking the limit
$N\rightarrow\infty$, $u\rightarrow 0$, with $Nu=\lambda$ fixed; we
get
\begin{eqnarray}
g_{Dd}(\{b_i\},\lambda) & = & \lim_{N u = \lambda} \left[ N\Gamma
_{Dd}^{(4)}(0,\{b_i\},u)\right] \nonumber \\
 & = & \frac{\lambda
}{1+\lambda \, \Pi_{Dd}(\{b_i\})}  .  \label{ECC}
\end{eqnarray}
It is clear that, while $g_{Dd}(\{b_i\},\lambda)$ depends on the
value of the fixed coupling constant $\lambda$ in a direct way, its
dependence on the compactifiation lengths is dictated by the
behavior of $\Pi_{Dd}$ as $\{b_i\}$ is varied. The dependence of
$g_{Dd}$ on $\{ L_i \}$ and $\lambda$ is the main point to be
discussed in the subsequent analysis.

Limiting behaviors of the finite coupling constant $g_{Dd}$ can be
readily obtained from the fact that $\Pi_{Dd}$ depends on $\{b_i\}$
through the Bessel functions of the third kind appearing in the
definition of the functions $F_{Dj}$, Eq.~(\ref{FDj}). First, if we
let all the compactification lengths tend to infinity, that is $\{
b_i\rightarrow 0 \}$, thus reducing the problem to the free space at
$T=0$, then $\Pi_{Dd}\rightarrow 0$ and we obtain, consistently,
that
\begin{equation}
\lim_{\{L_i\rightarrow\infty\}} g_{Dd}(\{b_i\},\lambda)=\lambda  ,
\end{equation}
where $\lambda$ is the fixed coupling constant in free space at zero
temperature. This is a consequence of the fact that
$K_{\nu}(z\rightarrow \infty)\rightarrow 0$, for $\nu$ integer or
half-integer. In the opposite limit, for any $b_i$ tending to
$\infty$ (that is, if any compactification length $L_i$ goes to
$0$), the single bubble diagram $\Pi_{Dd}\rightarrow\infty$, since
$K_{\nu}(z)\rightarrow 0$ as $z\rightarrow\infty$. This implies that
the effective coupling constant $g_{Dd}$ vanishes, irrespective of
the value of $\lambda$, suggesting that the system presents an
asymptotic-freedom type of behavior for short distances and/or for
high temperatures.

From the extreme limits considered above two situations may emerge,
as one changes the compactification lengths from $0$ to $\infty$:
either $\Pi_{Dd}$ varies from $\infty$ to $0$ through positive
values, or $\Pi_{Dd}$ reaches $0$ before tending to $0$ through
negative values. The latter case, which may actually happen, would
lead to an interesting situation where a divergence of the effective
coupling constant would appear at finite values of the lengths
$L_i$. This possibility, and its consequences, will be investigated
explicitly in the following subsections, considering the
compactified GN model at $T=0$ for space-time dimensions $D=2,3,4$.
The discussion of finite temperature effects is postponed to
Sec.~\ref{Sec4}.

\subsection{$2$-$D$ compactified GN model at $T=0$}

For $D=2$ and $d=1$ (two-dimensional space-time with the spatial
coordinate compactified), we put $b_1=(mL)^{-2}$ in
Eqs.~(\ref{WDd},\ref{FDj}), then Eq.~(\ref{SigmaR}) becomes
\begin{equation}
\Pi_{21}(L) = 2\, E_1(2 m L) - E_1(m L) , \label{S21}
\end{equation}
where the function $E_1(x)$ is defined by
\begin{equation}
E_1(x) = \frac{1}{\pi} \sum_{n=1}^{\infty} \left[ - K_{0}(x n) + (x
n)\, K_{1}(x n) \right] . \label{E1}
\end{equation}
Notice that $\Pi$ is dimensionless for $D=2$.

We can calculate $\Pi_{21}(L)$ numerically by truncating the series
appearing in the definition of the function $E_1(y)$,
Eq.~(\ref{E1}), at some value $n=M$. For moderate and large values
of $mL$ (say, $mL\gtrsim 0.5$), one can take for $M$ a relatively
small value; e.g., for $mL=0.5$, choosing $M=36$ already leads to
the correct value of $\Pi_{21}$ to six decimal places. As $mL$
increases, the value of $M$ can be made smaller to give the same
precision. However, since the functions $K_0(z)$ and $K_1(z)$
diverge for $z\rightarrow 0$ and the summation involves positive and
negative parcels, the calculation of $\Pi_{21}$ for small values of
$mL$ requires large values of $M$; for $mL = 0.005$, we need
$M\approx 4500$ to obtain $\Pi_{21}$ to six decimal places.
Fortunately, the relevant behavior of $\Pi_{21}(L)$ appears for
moderate values of $mL$ so that the numerical calculations are
carried out in a short computational time with very good precision.

\begin{figure}[ht]
\begin{center}
\scalebox{0.66}{{\includegraphics{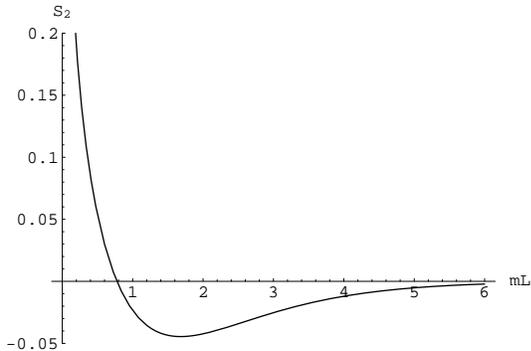}}}
\end{center}
\caption{Plot of $S_2 = \Pi_{21}(L)$ as a function of $mL$.}
\label{figS21}
\end{figure}

The function $\Pi_{21}(L)$ is plotted as a function of $mL$ in
Fig.~\ref{figS21}. From this figure and the numerical treatment of
Eq.~(\ref{S21}), we infer that $\Pi_{21}(L)$ diverges ($\rightarrow
+\infty$) when $L\rightarrow 0$ and tends to $0$, through negative
values, as $L\rightarrow\infty$. Also, we find that $\Pi_{21}(L)$
vanishes for a specific value of $L$, which we denote by $L_{{\rm
min}}^{(2)}$, being negative for all $L
> L_{{\rm min}}^{(2)}$, and assumes a minimum (negative) value
at a value of $L$ denoted by $L_{{\rm max}}^{(2)}$, for reasons that
will be clarified later. Numerically, it is found that $L_{{\rm
min}}^{(2)} \simeq 0.78\, m^{-1}\,$, $L_{{\rm max}}^{(2)} \simeq
1.68\, m^{-1}\,$ and $\Pi_{21}^{{\rm min}} \simeq -0.0445$. This
behavior of $\Pi_{21}$ as $L$ changes, particularly the fact that
$\Pi_{21}(L) < 0$ for $L > L_{{\rm min}}^{(2)}$, leads to remarkable
properties of the large-$N$ coupling constant $g_{21}(L,\lambda)$.
It is important to point out that such a dependence of the
polarization on $L$ is a direct consequence of the use of
anti-periodic boundary conditions for the spatial compactification.
Taking periodic boundary conditions (PBC), one would obtain
$\Pi_{21}^{{\rm PBC}}(L) = E_1(m L)/4$ which is positive for all
values of $L$, and so no significant size-effect would exist.

\begin{figure}[ht]
\begin{center}
\scalebox{0.66}{{\includegraphics{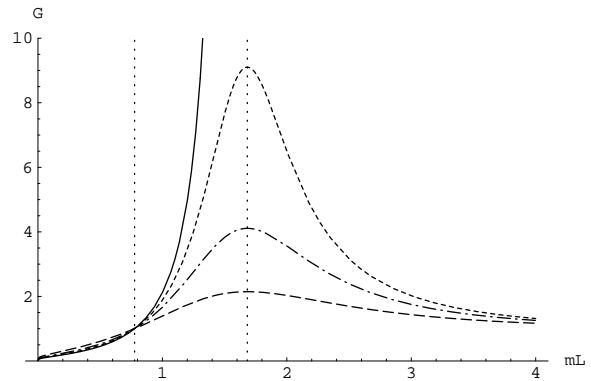}}}
\end{center}
\caption{Plots of the relative effective coupling constant,
$G=g_{21}(L,\lambda)/\lambda$, as a function of $mL$ for some values
of $\lambda$: $12.0$ (dashed line), $17.0$ (dotted-dashed line),
$20.0$ (dotted line) and $22.5$ (full line). The dotted vertical
lines, passing by $L_{\rm min}^{(2)} \simeq 0.78\, m^{-1}$ and
$L_{\rm max}^{(2)}\simeq 1.68\, m^{-1}$, are plotted as a visual
guide.} \label{figG21}
\end{figure}

Recall that, in the present case, Eq.~(\ref{ECC}) becomes
\begin{equation}
g_{21}(L,\lambda) = \frac{\lambda}{1+\lambda\, \Pi_{21}(L)} .
\label{G21}
\end{equation}
The divergence of $\Pi_{21}(L)$ as $L\rightarrow 0$ ensures that,
independently of the value of $\lambda$, $g_{21}(L,\lambda)$
approaches $0$ in this limit and, therefore, the system presents a
kind of asymptotic-freedom behavior for short distances. On the
other hand, since $\Pi_{21}(L)$ assumes negative values for $L >
L_{{\rm min}}^{(2)}$, the denominator of Eq.~(\ref{G21}) will
vanishe at a finite value of $L$ if $\lambda$ is sufficiently high.
This means that, starting from a low value of $L$ (within the region
of asymptotic freedom) and increasing the size of the system,
$g_{21}$ will diverge at a finite value of $L$,
$L_{c}^{(2)}(\lambda)$, if $\lambda$ is greater than the ``critical
value" $\lambda_c^{(2)} = (- \Pi_{21}^{{\rm min}})^{-1} \simeq
22.47$. We interpret this result by stating that, in the
strong-coupling regime ($\lambda \geq \lambda_c^{(2)}$) the system
gets spatially confined in a segment of length
$L_{c}^{(2)}(\lambda)$. The behavior of the $L$-dependent coupling
constant as a function of $mL$ is illustrated in Fig.~\ref{figG21},
for some values of the fixed coupling constant $\lambda$.

It should be emphasized that we are treating the massive GN model
with an arbitrary (but fixed) fermion mass. In this case, the model
does not possesses chiral symmetry which is explicitly broken. Since
we do not expect that this symmetry appears beyond the critical
value (with the radiatively corrected mass vanishing identically),
the instability indicated by the divergence of the coupling constant
is interpreted as signaling a spatial confining transition.

For $\lambda = \lambda_c^{(2)}$, by definition, the divergence of
$g_{21}(L,\lambda)$ is reached as $L$ approaches the value that
makes $\Pi_{21}$ minimal, which we have denoted by $L_{{\rm
max}}^{(2)}$. In the other limit, since
$g_{21}^{-1}(L,\lambda\rightarrow \infty) = \Pi_{21}(L)$,
$L_{c}^{(2)}(\lambda)$ tends to $L_{{\rm min}}^{(2)}$, the zero of
$\Pi_{21}(L)$, as $\lambda\rightarrow \infty$. In other words, the
confining length $L_{c}^{(2)}(\lambda)$ decreases from the maximum
value $L_{{\rm max}}^{(2)}$, when $\lambda = \lambda_c^{(2)}$,
tending to the lower bound $L_{{\rm min}}^{(2)}$ in the limit
$\lambda \rightarrow \infty$. The behavior of $L_{c}^{(2)}$, as a
function of $\lambda$, will be presented later.

\subsection{Compactified $3$-$D$ GN model at $T=0$}

For the $3$-$D$ model at zero temperature with two compactified
dimensions ($d=2$), denoting the compactification lengths associated
with the two spatial coordinates $x_1$ and $x_2$ by $L_1$ and $L_2$
($m^{-1} / \sqrt{b_1}$ and $m^{-1} / \sqrt{b_2}$, respectively),
formulas in Eq.~(\ref{WDd}) and Eq.~(\ref{SigmaR}) give
\begin{eqnarray}
\Pi_{32}(b_1,b_2) & = & \frac{m}{\sqrt{2}\pi^{\frac{3}{2}}}
\left[ 2 F_{31}(2;b_1) - F_{31}(2;4b_1)  \right. \nonumber \\
 & & +\, 2 F_{31}(2;b_2) - F_{31}(2;4b_2) - 2 F_{31}(1;b_2)
\nonumber \\
 & &  -\, 2 F_{31}(1;b_1) + F_{31}(1;4b_1) + F_{31}(1;4b_2) \nonumber
 \\
 & &  +\, 4 F_{32}(2;b_1,b_2) - 2 F_{32}(2;4b_1,b_2) \nonumber \\
 & &  -\, 2 F_{32}(2;b_1,4b_2) + F_{32}(2;4b_1,4b_2)
 \nonumber \\
  & &  -\, 4 F_{32}(1;b_1,b_2) + 2 F_{32}(1;4b_1,b_2)
  \nonumber \\
& & + \left. 2 F_{32}(1;b_1,4b_2) - F_{32}(1;4b_1,4b_2) \right]  .
\label{S32b1b2}
\end{eqnarray}
The functions $F_{3j}$ ($j=1,2$), specified in Eq.~(\ref{FDj}),
involve the Bessel functions of order $\pm\frac{1}{2}$, which are
expressed in terms of elementary functions:
\begin{equation}
K_{\pm\frac{1}{2}}(z) = \sqrt{\pi}\, \frac{\exp(-z)}{\sqrt{2z}}  .
\label{K}
\end{equation}
Thus, the series defining the functions $F_{3j}$, for both $j=1,2$,
are geometric series which can be summed up. Using Eq.~(\ref{K}) in
the expression of Eq.~(\ref{FDj}) and replacing $b_i$ by $L_i^{-2}$
(which corresponds to taking all the compactification lengths
measured in units of $m^{-1}$, as will be done from now on), we
obtain
\begin{eqnarray}
\Pi_{32}(L_1,L_2) & = & \frac{m}{2\pi} \left[ \frac{1} {L_1}\log(1 +
e^{-L_1}) - \frac{1}{1 + e^{L_1}}
 \right.
\nonumber \\
  & &  + \left. \frac{1}{L_2}\log(1 + e^{-L_2})  +
\frac{1}{1 + e^{L_2}} \right] \nonumber \\
 & & +\, \frac{m}{\pi} \left[ G_2(L_1,L_2) - 2\,
G_2(L_1,2L_2) \right. \nonumber \\
 & & - \left. 2\, G_2(2L_1,L_2) + 4\,
G_2(2L_1,2L_2) \right] , \nonumber \\
 & & \label{Sigma32N}
\end{eqnarray}
where the function $G_2(x,y)$ is defined by
\begin{eqnarray}
G_2(x,y) & = & \sum_{n,l=1}^{\infty} \exp\left( - \sqrt{x^2 n^2 +
y^2 l^2} \right) \nonumber \\
 & & \times \left[ 1 - \frac{1}{\sqrt{x^2 n^2 +
y^2 l^2}} \right] . \label{G2}
\end{eqnarray}
The numerical computation of $\Pi_{32}(L_1,L_2)$ is greatly
facilitated by the fact that the double series defining the function
$G_2(y,z)$ is rapidly convergent. The need of truncating the
summations at a larger value $n=M$ when $L_1$ and $L_2$ are very
small, as in the case of $D=2$, still exists but to a much less
extent.

\begin{figure}[ht]
\begin{center}
\scalebox{0.66}{{\includegraphics{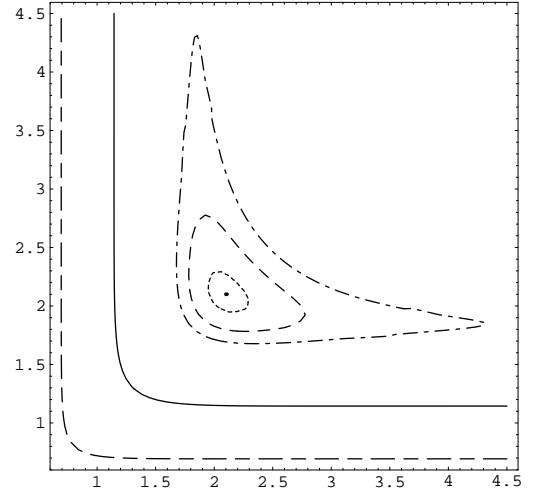}}}
\end{center}
\caption{Contour plots of $\Pi_{32}(L_1,L_2)/m$, with $L_1$ and
$L_2$ in units of $m^{-1}$. The open dashed line corresponds to
$\Pi_{32}(L_1,L_2)/m = 0.04$, the full line gives the points where
$\Pi_{32} = 0$, while the closed curves are for negative values of
$\Pi_{32}/m$, $-0.0091$, $-0.0095$ and $-0.0098$ (dashed-dotted,
dashed and dotted lines, respectively). The dot is the location of
the absolute minimum of $\Pi_{32}(L_1,L_2)$, which occurs for
$L_1=L_2\simeq 2.1\, m^{-1}$.} \label{ScontourLL}
\end{figure}

It is to be noticed that if either compactification length $L_1$ or
$L_2$ tends to $\infty$, all terms depending on it disappears from
Eq.~(\ref{Sigma32N}) and we regain the finite bubble diagram for the
case where only one spatial dimension is compactified in the $3$-$D$
model \cite{GNN}. Now, if both $L_1$ and $L_2$ tend simultaneously
to $\infty$, $\Pi_{32}$ goes to zero and $g_{32} \rightarrow
\lambda$, as expected. On the other hand, if either $L_1$ or $L_2$
tends to $0$, $\Pi_{32}\rightarrow +\infty$ implying that the system
gets asymptotically free, with the effective coupling constant
vanishing in this limit. The overall behavior of the bubble diagram
is illustrated in Fig.~\ref{ScontourLL}, where we draw the contour
plots of $\Pi_{32}(L_1,L_2)/m$. The full line in
Fig.~\ref{ScontourLL} is the locus of the points such that
$\Pi_{32}(L_1,L_2)/m = 0$, which for large $L_1$ ($L_2$) approaches
the straight line $L_1 = 1.14\, m^{-1}$ ($L_2 = 1.14\, m^{-1}$);
$\Pi_{32}(L_1,L_2)$ is positive below this curve, negative above it,
and reaches an absolute minimum, $\Pi_{32}^{{\rm min}} \simeq
-0.00986\, m$, at the point $L_1 = L_2 \simeq 2.10\, m^{-1}$.

\begin{figure}[ht]
\begin{center}
\scalebox{0.66}{{\includegraphics{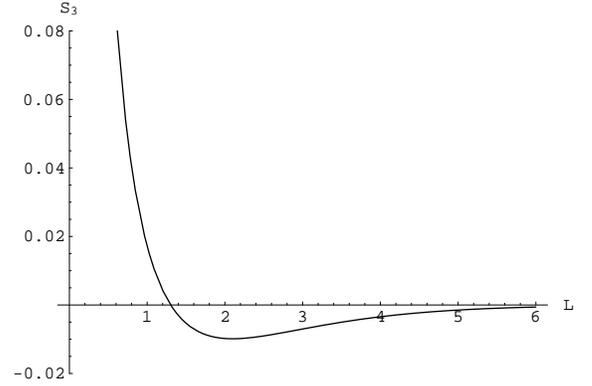}}}
\end{center}
\caption{Plot of $S_3 = \Pi_{32}(L)/m$ as a function of $L$, in
units of $m^{-1}$.} \label{S3LL}
\end{figure}

The fact that $\Pi_{32}$ assumes negative values in the whole region
of the parameter space $(L_1,L_2)$ above the full line in
Fig.~\ref{ScontourLL} implies that, for large enough values of
$\lambda$, $g_{32}$ will diverge at finite values of $L_i$, $i=1,2$.
However, to avoid unnecessary complication, our analysis is
restricted to the case where the system is confined within a square
of size $L$, by considering $L_1=L_2=L$. In other words, we shall
concentrate on the behavior of $\Pi_{32}$ along the diagonal of
Fig.~\ref{ScontourLL}. We plot, in Fig.~\ref{S3LL},
$S_3(L)=\Pi_{32}(L)/m$ as a function of $L$. We find that $S(L)$
vanishes for a specific value of $L$, $L_{{\rm min}}^{(3)}$, being
negative for all $L\geq L_{{\rm min}}^{(3)}$. Also, $\Pi_{32}(L)$
reaches an absolute minimum (negative) value for a value of $L$ we
denote by $L_{{\rm max}}^{(3)}$. We find, numerically, that $L_{{\rm
min}}^{(3)} \simeq 1.30\, m^{-1}\,$ and $L_{{\rm max}}^{(3)} \simeq
2.10\, m^{-1}\,$, with $\Pi_{32}^{{\rm min}} \simeq -0.00986\, m$,
as stated before. This behavior of $\Pi_{32}(L)$ has profound
implications on the effective coupling constant.

With $D=3$ and $d=2$, Eq.~(\ref{ECC}) is rewritten as
\begin{equation}
g_{32}(L,\lambda) = \frac{\lambda}{1+\lambda\, \Pi_{32}(L)} ,
\label{G32}
\end{equation}
and we find that, for $\lambda\geq\lambda_c^{(3)}=(-\Pi_{32}^{{\rm
min}})^{-1}\simeq 101.42\, m^{-1}$, the denominator in
Eq.~(\ref{G32}) will vanish for a finite value of $L$,
$L_c^{(3)}(\lambda)$, leading to a divergence in the effective
coupling constant. The behavior of the effective coupling constant
as a function of $L$, for increasing values of the fixed coupling
constant $\lambda$, can be illustrated showing the same pattern as
that of Fig.~\ref{figG21} for the preceding case. Similarly, we find
that the divergence occurs at $L_c^{(3)}(\lambda)$ which satisfies
$L_{{\rm min}}^{(3)} < L_c^{(3)}(\lambda) \leq L_{{\rm max}}^{(3)}$.
Again, we interpret such a result by considering the system
spatially confined in the sense that, starting with small $L$ (in
the region of asymptotic freedom), the size of the square cannot
increase above $L_c^{(3)}(\lambda)$, since
$g_{32}(L,\lambda)\rightarrow\infty$ as $L\rightarrow
L_c^{(3)}(\lambda)$.

\subsection{$D=4$ case at zero temperature}

In the case of the $4$-$D$ GN model with all three spatial
coordinates compactified, replacing $b_{i}$ by $L_{i}^{-2}$ (again,
$L_{i}$ measured in units of $m^{-1}$) into Eqs.~(\ref{WDd}),
(\ref{FDj}) and (\ref{SigmaR}) gives
\begin{eqnarray}
\Pi_{43}(\{L_i\}) & = & m^2 \left[ 2 H_1(2L_1) + 2 H_1(2L_2) + 2
H_1(2L_3)  \right.
\nonumber \\
 & & -\, H_1(L_1) - H_1(L_2) - H_1(L_3) \nonumber \\
 & & +\, 2 H_2(L_1,L_2) + 2 H_2(L_1,L_3)
  \nonumber \\
 & & +\, 2 H_2(L_2,L_3) -  4  H_2(L_1,2L_2) \nonumber \\
 & & -\, 4 H_2(L_1,2L_3) - 4 H_2(2L_1,L_2)
 \nonumber \\
 & &  -\, 4 H_2(2L_1,L_3) - 4  H_2(L_2,2L_3)
 \nonumber \\
 & & -\, 4 H_2(2L_2,L_3) + 8 H_2(2L_1,2L_2) \nonumber \\
 & & +\, 8 H_2(2L_1,2L_3) + 8  H_2(2L_2,2L_3)
  \nonumber \\
 & & -\, 4 H_3(L_1,L_2,L_3) + 8 H_3(2L_1,L_2,L_3)
 \nonumber \\
 & & +\, 8 H_3(L_1,2L_2,L_3) + 8 H_3(L_1,L_2,2L_3) \nonumber \\
 & & -\, 16 H_3(2L_1,2L_2,L_3) \nonumber \\
 & & -\, 16 H_3(2L_1,L_2,2L_3)
  \nonumber \\
 & &  -\, 16 H_3(L_1,2L_2,2L_3) \nonumber \\
 & & \left. +\, 32 H_3(2L_1,2L_2,2L_3)\right]  , \label{Sig43}
\end{eqnarray}
where the functions $H_j$, $j=1,2,3$, are defined by
\begin{eqnarray}
H_1(x) & = & \frac{1}{2\pi^2} \sum_{n=1}^{\infty} \left[ K_{0}(x n)
- \frac{K_{1}(x n)}{(x n)} \right] \, , \label{H1} \\
H_2(x,y) & = & \frac{1}{2\pi^2} \sum_{n,l=1}^{\infty} \left[ K_{0}
\left(\sqrt{x^2 n^2 + y^2 l^2} \right) \right. \nonumber \\
 & & -\left. \frac{K_{1}(\sqrt{x^2 n^2 + y^2 l^2})}{(\sqrt{x^2 n^2 + y^2
l^2})} \right] \, ,
\label{H2} \\
H_3(x,y,z) & = & \frac{1}{2\pi^2} \sum_{n,l,r=1}^{\infty} \left[
K_{0} \left(\sqrt{x^2 n^2 + y^2 l^2 + z^2 r^2} \right) \right.
\nonumber \\
 & & \left.
 -\, \frac{K_{1}(\sqrt{x^2 n^2 + y^2 l^2})}{(\sqrt{x^2
n^2 + y^2 l^2 + z^2 r^2})} \right]  . \label{H3}
\end{eqnarray}

Results for one or two compactified dimensions are obtained from
Eq.~(\ref{Sig43}) if two or one of the compactification lengths
become infinite. For example, taking $L_{2},L_{3} \rightarrow
\infty$, we get $\Pi_{41}(L_1) = m^2 [2H_1(2L_1) - H_1(L_1)]$, which
assumes the minimum value $\Pi_{41}^{{\rm min}} \simeq -0.001866\,
m^{2}$ at $L = L_{{\rm max}}^{(41)} \simeq 2.01\, m^{-1}$, leading
to the critical value $\lambda_{c}^{(41)} \simeq 535.91\, m^{-2}$,
and vanishes at $L = L_{{\rm min}}^{(41)} \simeq 1.43\, m^{-1}$. The
analysis made in Ref.~\cite{GNN} can be extended to this case of the
$4$-dimensional space with only one spatial coordinate compactified.
Similarly, we could discuss the situation with two compactified
spatial dimensions.

\begin{figure}[ht]
\begin{center}
\scalebox{0.72}{{\includegraphics{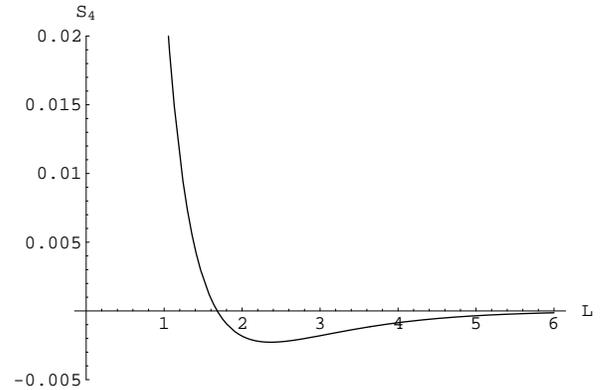}}}
\end{center}
\caption{Plot of $S_4 = \Pi_{43}(L)/m^2$ as a function of $L$, in
units of $m^{-1}$.} \label{Sigm43}
\end{figure}

Here, instead of dealing with all possibilities, we concentrate on
the case where the system is confined to a cubic box, that is, we
take $L_1=L_2=L_3=L$. With equal compactification lengths,
Eq.~(\ref{Sig43}) becomes
\begin{eqnarray}
\Pi_{43}(L) & = & m^2 \left[ 6 H_1(2L) - 3 H_1(L) + 6
H_2(L,L) \right. \nonumber \\
 & & -\, 24 H_2(L,2L) + 24 H_2(2L,2L) \nonumber \\
 & & -\, 4 H_3(L,L,L) + 24 H_3(L,L,2L) \nonumber \\
 & & \left. -\, 48 H_3(L,2L,2L) + 32 H_3(2L,2L,2L)
 \right] . \nonumber \\
 & & \label{Sigma43}
\end{eqnarray}
This quantity is plotted in Fig.~\ref{Sigm43}, which shows that it
has the same behavior as its counterparts for $D=2$ and $D=3$. We
find numerically that $\Pi_{43}(L)$ vanishes for $L = L_{\rm
min}^{(4)} \simeq 1.68\, m^{-1}$, being negative for $L > L_{\rm
min}^{(4)}$, and assumes the minimum value, $\Pi_{43}^{{\rm min}}
\simeq -0.0022751\, m^{2}$, when $L = L_{\rm max}^{(4)} \simeq
2.37\, m^{-1}$.

As in the other cases, the large-$N$ coupling constant,
\begin{equation}
g_{43}(L,\lambda) = \frac{\lambda}{1 + \lambda \Pi_{43}(L)} \, ,
\label{G43}
\end{equation}
diverges at a finite value of $L$, $L_c^{(4)}(\lambda)$, if $\lambda
\geq \lambda_c^{(4)} = -(\Pi_{43}^{{\rm min}})^{-1} \simeq 439.54\,
m^{-2}$, meaning that the system gets confined in a cubic box of
edge $L_c^{(4)}(\lambda)$ which is bounded in the interval between
the values $L_{\rm min}^{(4)}$ and $L_{\rm max}^{(4)}$.

\subsection{Dependence of $L_{c}^{(D)}$ on $\lambda$}

For the cases we have analyzed above, namely $D=2,3,4$ with all
spatial coordinates compactified and $L_1 = \cdots = L_{D-1} = L$,
we find that the confining length $L_{c}^{(D)}(\lambda)$ lies in a
finite interval,
\begin{equation}
L_{c}^{(D)}(\lambda) \in \left( L_{{\rm min}}^{(D)},L_{{\rm
max}}^{(D)} \right] , \label{intervalL}
\end{equation}
where the maximum value corresponds to $\lambda_{c}^{(D)}$, the
minimum value of the fixed coupling constant $\lambda$ allowing
spatial confinement, while $L_{{\rm min}}^{(D)}$ sets the bound as
$\lambda \rightarrow \infty$. Then, a question emerges of how
$L_{c}^{(D)}(\lambda)$ changes as $\lambda$ increases from
$\lambda_{c}^{(D)}$ to infinity.

\begin{figure}[ht]
\begin{center}
\scalebox{0.66}{{\includegraphics{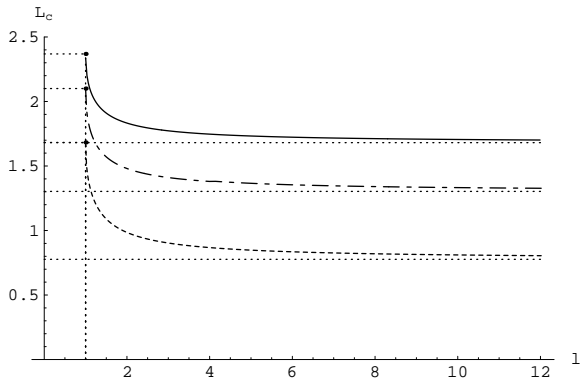}}}
\end{center}
\caption{Plot of the confining length (in units of $m^{-1}$), as a
function of $l = \lambda/\lambda_c^{(D)}$, for $D=2,3,4$ (dashed,
dashed-dotted and full lines respectively); the horizontal dotted
lines correspond to the limiting values $L_{{\rm min}}^{(D)}$ and
$L_{\rm max}^{(D)}$ (given in the text), plotted as a visual guide.}
\label{LcLambdaD}
\end{figure}

For a given value of $\lambda$ ($\geq\lambda_{c}^{(D)}$), the
confining length $L_{c}^{(D)}(\lambda)$ can be found numerically by
determining the smallest root of the equation
\begin{equation}
g_{D D-1}^{-1}(L,\lambda) = \frac{1}{\lambda} \left[ 1 + \lambda
\Pi_{D D-1}(L) \right] = 0 .
\end{equation}
That is, following the interpretation provided before, starting from
small values of $L$, the first value at which $g_{D D-1}^{-1}$
vanishes does provide the confining length of the system,
$L_{c}^{(D)}(\lambda)$. In Fig.~\ref{LcLambdaD}, we plot
$L_{c}^{(D)}(\lambda)$ as a function of $l=\lambda /
\lambda_{c}^{(D)}$, for the cases $D=2,3,4$.

\section{Compactified GN model at finite temperature
\label{Sec4}}

We shall now consider the effect of raising the temperature on the
effective coupling constant for the GN model with all spatial
dimensions compactified. Finite-temperature effects are introduced
through the compactification of the time coordinate, with the
compactification ``length" given by $L_D = \beta = 1/T$, where $T$
is the temperature. It is important to emphasize that, although in
an Euclidean theory time and space coordinates are treated on the
same footing, the interpretation of their compactifications are
rather distinct; while compactification of spatial dimensions can be
thought as describing confined fields, time compactification
corresponds to taking the system in thermal equilibrium at
temperature $\beta^{-1}$.

We generally expect that the dependence of $\Pi_{D D}$ and $g_{D D}$
on $\beta$ should follow similar patterns as that for the dependence
on $L$. Like in the case where any compactification length $L_i$
tends to zero, we find that $\Pi_{D D}(L,\beta) \rightarrow \infty$
as $\beta \rightarrow 0$ ($T \rightarrow \infty$), implying that
$g_{D D} \rightarrow 0$ independently of the value of the fixed
coupling constant $\lambda$. This means that we have an
asymptotic-freedom behavior for very high temperatures.

For $\beta\rightarrow\infty$ ($T\rightarrow 0$), the behavior of
$\Pi_{D D-1}(L)$ has been described earlier: for sufficiently high
values of $\lambda$, the system is confined in a $(D-1)$-dimensional
cube of edge $L_{c}^{(D)}$. Based on these observations, we expect
that, starting from the compactified model at $T=0$ with $\lambda
\geq \lambda_{c}^{(D)}$, raising the temperature will lead to the
suppression of the divergence of $g_{D D}$ and the consequent
spatial deconfinement of the system, at a specific value of the
temperature, $T_{d}^{(D)}$. The way such a ``deconfining" transition
occurs and the determination of the deconfining temperature for the
cases of $D=2,3,4$ are the points addressed in the next subsections.

\subsection{$2$-$D$ compactified GN model at $T\neq 0$}

To account for the effect of finite temperature on the $2$-$D$
compactified GN model \cite{AIP}, we take the second Euclidean
coordinate (the imaginary time, $x_2$) compactified in a length $L_2
= \beta = 1/T$. In this case, replacing $b_1 = L^{-2}$ and $b_2 =
\beta^{-2}$ ($L$ and $\beta$ measured in units of $m^{-1}$) into
Eqs.~(\ref{WDd}), (\ref{FDj}) and (\ref{SigmaR}), the $L$ and
$\beta$-dependent bubble diagram can be written as
\begin{eqnarray}
\Pi_{22}(L,\beta) & = & 2\, E_1(2 L) - E_1(L) + 2\, E_1(2
\beta) - E_1(\beta)  \nonumber \\
 & & +\,  2\, E_2(L,\beta) - 4\, E_2(2 L,\beta) \nonumber \\
 & & -\, 4\, E_2(L,2 \beta) + 8\, E_2(2 L,2 \beta) , \label{S22}
\end{eqnarray}
where the function $E_1(x)$ is given by Eq.~(\ref{E1}) and the
function $E_2(x,y)$ is defined by
\begin{eqnarray}
E_2(x,y) & = & \frac{1}{\pi} \sum_{n,l=1}^{\infty} \left[ -
K_0\left( \sqrt{x^2 n^2 + y^2 l^2} \right) \right. \nonumber \\
 & & + \left.  \left( \sqrt{x^2 n^2 +
y^2 l^2} \right)\, K_1\left( \sqrt{x^2 n^2 + y^2 l^2} \right)
\right] . \nonumber \\
& & \label{E2}
\end{eqnarray}

We first notice that, due to the behavior of the Bessel functions
$K_0(z)$ and $K_1(z)$, all $\beta$-dependent terms in
Eq.~(\ref{S22}) vanish in the limit $\beta\rightarrow\infty$ and so
$\Pi_{22}(L,\beta)$ reduces to the expression for zero temperature,
$\Pi_{21}(L)$. On the other hand, if $\beta \rightarrow 0$,
$\Pi_{22}(L,\beta) \rightarrow \infty$ and, independently of the
value of $\lambda$, the system becomes asymptotically free.
Therefore, we expect that, within the strong-coupling regime,
raising the temperature leads to the suppression of the divergence
of $g_{22}$ and the disappearance of the spatial confinement. In
other words, for a given value of $\lambda \geq \lambda_c^{(2)}$,
there exists a temperature, $T_d^{(2)}(\lambda)$, above which
$g_{22}$ has no divergence and the system is spatially deconfined.

\begin{figure}[ht]
\begin{center}
\scalebox{0.66}{{\includegraphics{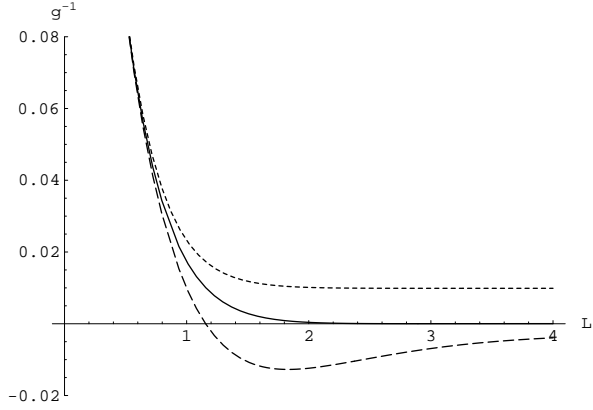}}}
\end{center}
\caption{Inverse of the effective coupling constant $g_{22}^{-1}$,
with $\lambda = 30$ fixed, as a function of  $L$ (in units of
$m^{-1}$), for some values of $\beta$ (in units of $m^{-1}$): $2.4$,
$1.15$ and $1.0$ (dashed, full and dotted lines, respectively).}
\label{gInv2}
\end{figure}

The deconfining temperature $T_d^{(2)}(\lambda)$ is determined by
analyzing the behavior of $g_{22}^{-1}(L,\beta,\lambda)$ as $T$ is
increased. This process is illustrated in Fig.~\ref{gInv2} where we
plot $g_{22}^{-1}(L,\beta,\lambda)$ as a function of $L$, for some
values of $\beta$ and a fixed value of $\lambda > \lambda_c^{(2)}$.
For this example with $\lambda = 30$, we find that the minimum value
of $g_{22}^{-1}$ vanishes for $\beta = \beta_d^{(2)} \simeq 1.15\,
m^{-1}$ and is positive for $\beta > \beta_d^{(2)}$. Thus, the
deconfining temperature, for $\lambda = 30$, is given by $T_d^{(2)}
= (\beta_d^{(2)})^{-1} \simeq 0.87\, m$. The full dependence of
deconfining temperature on $\lambda$ will be discussed later.

\subsection{$3$-$D$ compactified GN model at finite $T$}

We now take the time coordinate ($x_3$) compactified in a length
$\beta = 1/T$ to investigate the temperature effect in the $3$-$D$
compactified GN model. In Ref.~\cite{GNNT} the 3-$D$ model, with
only one spatial dimension compactified, was treated at finite
temperature \cite{BJP}. Here, we deal with the fully compactified
model. Taking $b_1 = b_2 = L^{-2}$ and fixing $b_3 = \beta^{-2}$ in
Eqs.~(\ref{WDd}), (\ref{FDj}) and (\ref{SigmaR}), the $L$-$\beta$
dependent bubble diagram is given by
\begin{eqnarray}
\Pi_{33}(L,\beta) & = & \frac{m}{2\pi} \left[ \frac{2} {L}\log(1 +
e^{-L}) - \frac{2}{1 + e^{L}}  \right. \nonumber \\
 & & + \left.\frac{1}{\beta}\log(1 + e^{-\beta}) -
 \frac{1}{1 + e^{\beta}} \right] \nonumber \\
 & & +\, \frac{m}{\pi} \left[ G_2(L,L) + 2 G_2(L,\beta)
 - 4 G_2(L,2L) \right. \nonumber \\
 & & -\, 4 G_2(2L,\beta) - 4 G_2(L,2\beta) + 4 G_2(2L,2L)  \nonumber \\
 & & +\, 8 G_2(2L,2\beta) - 2 G_3(L,L,\beta) \nonumber \\
   & &  +\, 4 G_3(L,L,2\beta) + 8 G_3(2L,L,\beta) \nonumber \\
 & & -\, 8 G_3(2L,2L,\beta) - 16 G_3(2L,L,2\beta) \nonumber \\
 & &  +\left. 16 G_3(2L,2L,2\beta) \right] ,  \label{Sigma33N}
\end{eqnarray}
where $G_2(x,y)$ is given by Eq.~(\ref{G2}) and the function
$G_3(x,y,z)$ is defined by
\begin{eqnarray}
G_3(x,y,z) & = & \sum_{n,l,r=1}^{\infty} \exp\left( - \sqrt{x^2 n^2
+ y^2 l^2 + z^2 r^2} \right) \nonumber \\
 & & \times \left[ 1 -
\frac{1}{\sqrt{x^2 n^2 + y^2 l^2 + z^2 r^2}} \right]  . \label{G3}
\end{eqnarray}
Notice that, taking $\beta\rightarrow\infty$, Eq.~(\ref{Sigma33N})
reduces to $\Pi_{32}(L)$, obtained from Eq.~(\ref{Sigma32N}) with
$L_1=L_2=L$.

As before, the increase of the temperature destroys the spatial
confinement that exists for $\lambda \geq \lambda_c^{(3)}$ at $T=0$.
We can determine the deconfining temperature by searching for the
value of $\beta(\lambda)$ for which the minimum of the inverse of
the effective coupling constant, $g_{33}^{-1}(L,\beta,\lambda) = (1
+ \lambda \Pi_{33}(L,\beta))/\lambda$, vanishes. For example, taking
the specific case of $\lambda = 110\, m^{-1}$, we find
$\beta_{d}^{(3)} \simeq 1.65\, m^{-1}$ which corresponds to the
deconfining temperature of $T_d^{(3)} \simeq 0.61\, m$; this result
can be illustrated in a figure with the same pattern as that
appearing in Fig.~\ref{gInv2} for the $D=2$ case.

\subsection{$D=4$ case at $T \neq 0$}

For the fully compactified $4$-$D$ GN model, fixing
$b_1=b_2=b_3=L^{-2}$ and $b_4 = \beta^{-2}$ ($L$ and $\beta $ in
units of $m^{-1}$), we find from Eqs.~(\ref{WDd}), (\ref{FDj}) and
(\ref{SigmaR}) that
\begin{eqnarray}
\Pi_{44}(L,\beta) & = & m^2 \left[ 6 H_1(2L) - 3 H_1(L) + 2
H_1(2\beta) \right. \nonumber \\
 & & -\, H_1(\beta) + 6 H_2(L,L) + 6 H_2(L,\beta) \nonumber \\
 & & -\, 24 H_2(L,2L) - 12 H_2(L,2\beta) \nonumber \\
 & & -\, 12 H_2(2L,\beta) + 24 H_2(2L,2L) \nonumber \\
 & & +\, 24 H_2(2L,2\beta) - 4 H_3(L,L,L) \nonumber \\
 & & -\, 12 H_3(L,L,\beta) + 24 H_3(L,L,2L) \nonumber \\
 & & +\, 48 H_3(L,2L,\beta) + 24 H_3(L,L,2\beta) \nonumber \\
 & & -\, 48 H_3(L,2L,2L) - 48 H_3(2L,2L,\beta) \nonumber \\
 & & -\, 96 H_3(L,2L,2\beta) + 32 H_3(2L,2L,2L) \nonumber \\
 & & +\, 96 H_3(2L,2L,2\beta) + 8 H_4(L,L,L,\beta)
 \nonumber \\
 & & -\, 48 H_4(L,L,2L,\beta) + 192 H_4(L,2L,2L,\beta) \nonumber \\
 & & -\, 16 H_4(L,L,L,2\beta) + 96 H_4(L,L,2L,2\beta) \nonumber \\
 & &  -\, 64 H_4(2L,2L,2L,\beta)  \nonumber \\
 & & -\, 192 H_4(L,2L,2L,2\beta)  \nonumber \\
 & & +\left. 128 H_4(2L,2L,2L,2\beta) \right]  , \label{Sigma43}
\end{eqnarray}
where the functions $H_1$, $H_2$ and $H_3$ are given by
Eqs.~(\ref{H1}-\ref{H3}), and $H_4(x,y,z,w)$ is defined by
\begin{widetext}
\begin{equation}
H_4(x,y,z,w) = \frac{1}{2\pi^2} \sum_{n,l,r,s=1}^{\infty} \left[
K_{0}\left( \sqrt{x^2 n^2 + y^2 l^2 + z^2 r^2 + w^2 s^2} \right) -
\frac{K_{1}\left( \sqrt{x^2 n^2 +
 y^2 l^2 +
z^2 r^2 + w^2 s^2} \right)}{\sqrt{x^2 n^2 + y^2 l^2 + z^2 r^2 + w^2
s^2}} \right] .
\end{equation}
\end{widetext}

Proceeding as before, we determine the deconfining temperature by
searching for the value of $\beta(\lambda)$ for which the minimum of
the inverse of the effective coupling constant,
$g_{44}^{-1}(L,\beta,\lambda) = (1 + \lambda
\Pi_{44}(L,\beta))/\lambda$, vanishes. For example, taking the
specific case of $\lambda = 620\, m^{-2}$, we find $\beta_{d}^{(4)}
\simeq 1.707\, m^{-1}$ which corresponds to the deconfining
temperature of $T_d^{(4)} \simeq 0.59\, m$.

\subsection{Dependence of $T_{d}^{(D)}$ on $\lambda$}

We now determine the dependence of the deconfining temperature on the
fixed coupling constant. For the system with all spatial dimensions
compactified and at finite temperature, we have
\begin{equation}
g_{DD}^{-1}(L,\beta;\lambda) = \frac{1}{\lambda} \left[ 1 + \lambda
\, \Pi_{DD}(L,\beta) \right] . \label{gLbeta}
\end{equation}
As discussed before, the system is deconfined at a given temperature
if the minimal value of $g_{DD}^{-1}(L,\beta;\lambda)$ with respect
to changes in $L$ is positive. It was also argued that, no matter
how high the value of $\lambda$ is, the system becomes deconfined
above a given temperature, $T_{d}^{(D)}(\lambda)$. In fact, one
expects that
\begin{equation}
T_{d}^{(D)}(\lambda) \in \left[ T_{\rm min}^{(D)},T_{\rm max}^{(D)}
\right) , \label{intervalT}
\end{equation}
with the limiting values corresponding to $\lambda_{c}^{(D)}$ and
$\lambda\rightarrow \infty$, respectively.

From Eq.~(\ref{gLbeta}), we find that $\,\min_{\{ L \}}
g_{DD}^{-1}(L,\beta;\lambda) = \left[ 1 + \lambda \, M_{D}(\beta)
\right] / \lambda$, where we have defined the function
\begin{equation}
M_{D}(\beta) = \min_{\{ L \}}\Pi_{DD}(L,\beta) .
\end{equation}
For a fixed value of $\lambda$, the behavior of the minimum value of
$g_{DD}^{-1}(L,\beta;\lambda)$ relative to changes in $L$ is
dictated by the function $M_{D}(\beta)$. Notice that,
$M_{D}(\beta\rightarrow\infty) = \Pi_{D D-1}^{R{\rm min}} = -[
\lambda_{c}^{(D)} ]^{-1}$, while $M_{D}(\beta) \rightarrow \infty$
as $\beta \rightarrow 0$.

\begin{figure}[ht]
\begin{center}
\scalebox{0.66}{{\includegraphics{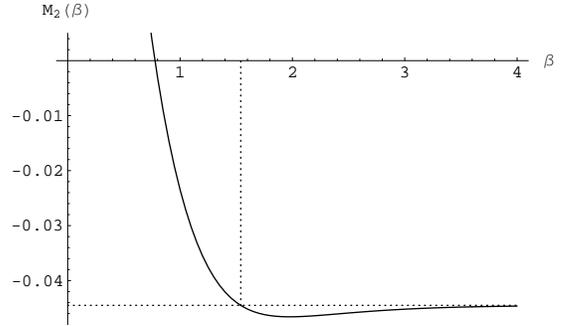}}}
\end{center}
\caption{Minimal values of $\Pi_{22}(L;\beta)$ with respect of
changes in $L$, as a function of $\beta$ (in units of $m^{-1}$). The
dotted horizontal line corresponds to the value $-[
\lambda_{c}^{(2)} ]^{-1}$.} \label{M2}
\end{figure}

Consider, initially, the case $D=2$. The function $M_{2}(\beta)$ is
illustrated in Fig.~\ref{M2}. We find that, as $\beta$ decreases
from $\infty$ (i.e., $T$ increases from $0$), the minimum value of
$\Pi_{22}(L,\beta)$ (with respect to changes in $L$) starts to
decrease from negative values, passes through the lowest value and
then starts to increase, reaching zero at a certain value of $\beta$
below which the minima of $\Pi_{22}$ are positive. Thus, for
$\lambda = \lambda_{c}^{(2)}$, increasing the temperature from zero,
the system remains confined until the temperature reaches the value
$T_{\rm min}^{(2)} = [\beta_{\rm max}^{(2)}]^{-1} \simeq 0.65\, m$,
corresponding to the finite solution of the equation $1 +
\lambda_{c}^{(2)} M_{2}(\beta)= 0$, $\beta = \beta_{\rm max}^{(2)}
\simeq 1.54\, m^{-1}$, which is indicated by the vertical dotted
line in Fig.~\ref{M2}. Now, if we take $\lambda >
\lambda_{c}^{(2)}$, the deconfining temperature is obtained from the
solution of the equation $1 + \lambda M_{2}(\beta)= 0$ which can be
determined from the intercept of the horizontal line at
$-\lambda^{-1}$ (lying above the line $-[ \lambda_{c}^{(2)} ]^{-1}$
and below the $\beta$ axis) and the graph of $M_2(\beta)$.
Naturally, as one takes $\lambda \rightarrow \infty$, the existence
of a solution of this equation requires $M_2(\beta) \rightarrow 0$;
this lower (open) limit occurs at the value $\beta = \beta_{\rm
min}^{(2)} \simeq 0.776\, m^{-1} $, corresponding to the temperature
$T_{\rm max}^{(2)} \simeq 1.29\, m$. The dependence of the
deconfining temperature $T_{d}^{(2)}$ on $\lambda$ is determined
numerically. We find that, for $\lambda$ not close to
$\lambda_{c}^{(2)}$ (that is, for $\lambda \gtrsim 1.5\,
\lambda_{c}^{(2)}$), $\beta_{d}^{(2)}(\lambda) =
L_{c}^{(2)}(\lambda)$ within six decimal places; this approximate
equality, valid for large values of $\lambda$, is a consequence of
the symmetry of the expression for $\Pi_{22}(L,\beta)$ by the change
$L \leftrightarrow \beta$. The whole behavior of
$T_{d}^{(2)}(\lambda)$ is shown in Fig.~\ref{Tdlamb}, together with
the other cases.

It should be remarked that the absolute minimum of
$\Pi_{22}(L,\beta)$, $\Pi_{22}^{{\rm min}} \simeq -0.0466$, is
slightly smaller than $\Pi_{21}^{{\rm min}}$ and occurs at the point
with $L = \beta \simeq 1.98\, m^{-1}$. This means, as shown in
Fig.~\ref{M2}, that the minimum value of $M_2(\beta)$ does not occur
at zero temperature. This leads to an anomalous situation, for a
small range of values of the coupling constant $\lambda$ ($21.46 <
\lambda < \lambda_c^{(2)} \simeq 22.47$), in which no singularity
exists at $T = 0$ but the equation $1 + \lambda M_2(\beta) = 0$
possesses two solutions for finite values of $\beta$; this would
imply that the system being unconfined at $T = 0$ would get confined
at a finite temperature and then becomes unconfined again at a
smaller value of $\beta$. Such a situation, which emerges from the
mathematical structure of the zeta-function regularization, has no
physical meaning and will be discarded; we shall only consider the
strong coupling regime which is free from pathologies.

For $D=3$, the graph of the function $M_{3}(\beta)/m$ has a form
similar to that of $M_{2}(\beta)$ (illustrated in Fig.~\ref{M2}) so
that the same reasoning leads to the deconfining temperature
$T_{d}^{(3)}(\lambda) = [\beta_{d}^{(3)}(\lambda)]^{-1}$, where
$\beta_{d}^{(3)}(\lambda)$ is the finite root of the equation $1 +
\lambda M_{3}(\beta) = 0$. In this case, the limiting values are
$\beta_{\rm max}^{(3)} \simeq 1.85\, m^{-1}$ and $\beta_{\rm
min}^{(3)} \simeq 1.14\, m^{-1}$, associated with $\lambda =
\lambda_{c}^{(3)}$ and $\lambda \rightarrow \infty$, corresponding
to $T_{\rm min}^{(3)} \simeq 0.54 m$ and $T_{\rm max}^{(3)} \simeq
0.88 m$. The plot of $T_{d}^{(3)}(\lambda)$, as a function of
$l=\lambda / \lambda_{c}^{(3)}$, is also shown in Fig.~\ref{Tdlamb}.
Distinctly from the $D=2$ case, where all the expressions are
symmetric by the change $L\leftrightarrow\beta$, here we find that
$\beta_{d}^{(3)}(\lambda) < L_{c}^{(3)}(\lambda)$ for all $\lambda
\geq \lambda_{c}^{(3)}$, the difference being of the order of
$10\%$. The maximum value of $T_{d}^{(3)}$ ($\simeq 0.87\, m$),
which occurs as $\lambda \rightarrow \infty$, corresponds to
$\beta_{d}^{(3)}$ reaching the value of the minimal confining length
($\simeq 1.14\, m^{-1}$) when only one coordinate is compactified.

\begin{figure}[ht]
\begin{center}
\scalebox{0.76}{{\includegraphics{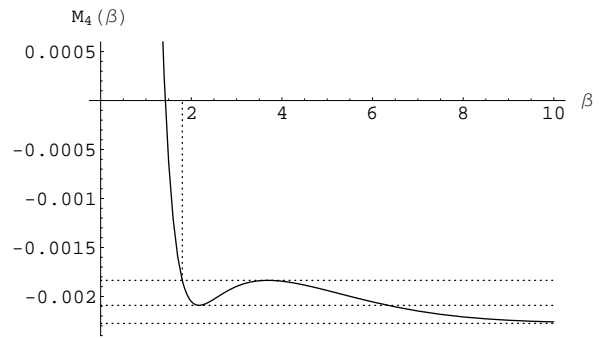}}}
\end{center}
\caption{Minimal values of $\Pi_{44}(L;\beta)/m^2$ with respect of
changes in $L$, as a function of $\beta$ (in units of $m^{-1}$). The
dotted horizontal lines correspond to the values $-[
\lambda_{c}^{(4)} ]^{-1} < M_{\rm min} < M_{\rm max}$.} \label{M4}
\end{figure}

\begin{figure}[ht]
\begin{center}
\scalebox{0.76}{{\includegraphics{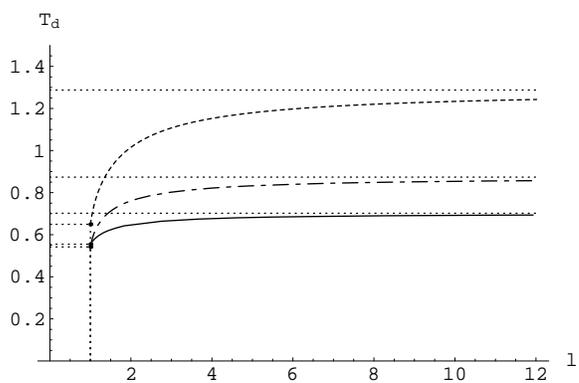}}}
\end{center}
\caption{Deconfining temperature $T_d^{(D)}(\lambda)$ (in units of
$m$), as a function of $l = \lambda/\lambda_c^{(D)}$, for $D=2,3,4$
(dashed, dashed-dotted and full lines respectively); the horizontal
dotted lines correspond to the limiting values $T_{{\rm min}}^{(D)}$
and $T_{\rm max}^{(D)}$ (given in the text), plotted as a visual
guide.} \label{Tdlamb}
\end{figure}

The $D = 4$ case is more subtle due to the behavior of the function
$M_{4}(\beta)$, which is presented in Fig.~\ref{M4}. We find that
the minimum value of $\Pi_{44}(L,\beta)$ (with respect to variations
of $L$) starts to increase from negative values as $\beta$ is
diminished from $\infty$, reaches a local maximum value ($M_{\rm
max} \simeq -0.0018356\, m^2$), decreases to a local minimum
($M_{\rm min} \simeq -0.002091\, m^2$) before increasing to reach
zero and become positive. Therefore, for $\lambda =
\lambda_{c}^{(4)} = - [M_4(\beta\rightarrow\infty)]^{-1} \simeq
439.54\, m^{-2}$, no finite solution of the equation $1 +
\lambda_{c}^{(4)} M_4(\beta) = 0$ exists, which means that the
system is deconfined if the temperature is greater than zero, no
matter how small it is; that is, for $\lambda = \lambda_{c}^{(4)}$,
spatial confinement is only possible strictly at $T = 0$. For
$\lambda_{c}^{(4)} < \lambda < 478.2\, m^{-2} $ ($\simeq - M_{\rm
min}^{-1}$), the equation $1 + \lambda M_4(\beta) = 0$ possesses one
solution occurring at finite $\beta$, which could eventually be
interpreted as leading to a deconfining temperature. However, if
$478.2\, m^{-2} < \lambda < 544.8\, m^{-2}$ ($\simeq - M_{\rm
max}^{-1}$), the equation $1 + \lambda M_4(\beta) = 0$ has three
distinct finite solutions. If we interpret the highest one as giving
the deconfining temperature, we would have to face the puzzling
situation in which the system would reenter a spatially confined
phase for $\beta$ ranging between the other two smaller solutions.
Such an anomalous behavior can be avoided if we redefine the strong
coupling regime by considering the range $-M_{\rm max}^{-1} <
\lambda < \infty$.

With such a redefinition of the strong coupling regime for $D = 4$,
which amounts to considering the lowest value of $\lambda$ leading
to spatial confinement as being $\lambda_{c}^{(4)} = - M_{\rm
max}^{-1} \simeq 544.8\, m^{-2}$, we find a well defined deconfining
temperature obtained from the intercept of the horizontal line at
$-\lambda^{-1}$ and the curve $M_4(\beta)$. We get $\beta_{\rm
max}^{(4)} \simeq 1.80\, m^{-1}$ (indicated by the vertical dotted
line in Fig.~\ref{M4}) and $\beta_{\rm min}^{(4)} \simeq 1.43\,
m^{-1}$ (where $M_4$ vanishes), corresponding to the limits of the
deconfining temperature $T_{\rm min}^{(4)} \simeq 0.55\, m$ and
$T_{\rm max}^{(4)} \simeq 0.70\, m$, respectively. Similar to the
$D=3$ case, the value of $\beta_{\rm min}^{(4)}$, giving the upper
bound for $T_{d}^{(4)}$ ($\lambda\rightarrow\infty$), is identical
to the smallest confining length when only one spatial dimension is
compactified. The overall behavior of $T_{d}^{(4)}(\lambda)$, found
numerically, is presented in Fig.~\ref{Tdlamb} together with the
cases of $D=2$ and $3$.

\section{Concluding remarks}

We have used methods of quantum field theory in toroidal topologies
to investigate the behavior of the D-dimensional massive,
$N$-component, Gross-Neveu model with compactified spatial
dimensions. The model is treated both at zero and finite
temperatures. We calculate the large-$N$ coupling constant, $g_D$,
as a function of the compactification length $L$, the temperature
$T=\beta^{-1}$ and the fixed coupling constant $\lambda$, i.e. $g_D
= g_D(L,\beta ;\lambda)$. We find that, for either $L\rightarrow 0$
or $T\rightarrow\infty$, irrespective of the value of $\lambda$,
$g_D$ tends to $0$ indicating that the system presents a sort of
asymptotic-freedom behavior in these limits, where the effective
interaction between the fermions vanishes.

For $T=0$, in the strong-coupling regime ($\lambda \geq
\lambda_{c}^{(D)}$), increasing $L$ from low values (within the
asymptotic freedom region) leads to a divergence of $g_D$ at a
finite critical value $L=L_{c}^{(D)}(\lambda)$, signaling the
existence of a second-order phase transition, as suggested by the
linear response theory. Also, since we consider a four-fermion
interaction model at zero chemical potential, we should not expect a
first-order phase transition for any value of the parameters
characterizing the system, as happens for massless models. We
interpret this singularity as indicating that the system gets
spatially confined in a $(D-1)$-dimensional box of edge
$L_{c}^{(D)}(\lambda)$. We have shown that $L_{c}^{(D)}(\lambda) =
f_D(\lambda)\, m^{-1}$, where the functions $f_D(\lambda)$ are
plotted in Fig.~\ref{LcLambdaD}, for $D=2,3,4$.

As $T$ is raised from $0$, with $\lambda \geq \lambda_{c}^{(D)}$,
the minimum value of $g_{D}^{-1}(L,\beta;\lambda)$, with respect to
changes in $L$ for fixed values of $\beta$, increases from negative
values reaching $0$ at a temperature $T_{d}^{(D)}(\lambda)$ above
which $g_D$ does not present any divergence. We interpret this fact
as the system being deconfined for temperatures higher than
$T_{d}^{(D)}(\lambda)$. To avoid any anomalous behavior, we have
redefined the value of $\lambda_{c}^{(4)}$ augmenting the lower
bound which defines the strong-coupling regime when $D=4$. In any
case, we find that $T_{d}^{(D)}(\lambda) = h_D(\lambda)\, m$, where
$h_D(\lambda)$ are the functions plotted in Fig.~\ref{Tdlamb}, for
$D=2,3,4$. It is to be noted that, for the redefined value
$\lambda_{c}^{(4)} \simeq 544.8\, m^{-2}$, the zero-temperature
maximum confining length is given by $L_{\rm max}^{(4)} \simeq
2.00\, m^{-1}$.

It is worth emphasizing that the dependencies of $L_{c}^{(D)}$ and
$T_{d}^{(D)}$ on the parameters $\lambda$ and $m$ are intrinsic
results of the model, that is, they do not emerge from any
adjustment. The dependence on $m$ is precisely what one expects from
dimensional arguments, with $L$ and $\beta$ being proportional to
$m^{-1}$ in natural units. But since the functions $f_D(\lambda)$
and $h_D(\lambda)$ take on values in finite intervals (in all cases,
contained in $[0.5,2.5]$), we find that extremely light fermions
($m\rightarrow 0$) are not confined at all, while extremely heavy
ones ($m\rightarrow \infty$) would be strictly confined in a dot, no
matter what the value of $\lambda$. Also, it should be noticed that
the product ${\cal
 P}_D(\lambda) = L_{c}^{(D)}(\lambda) \, T_{d}^{(D)}(\lambda) =
f_D(\lambda) \, h_D(\lambda)$ is dimensionless and very close to the
unit in the whole strong coupling regime, $\lambda \geq
\lambda_c^{(D)}$, irrespective of the value of $D$; actually, one
finds that ${\cal P}_D(\lambda) \in \left( 1.00,1.18 \right)$ for
all cases. Such an inverse relation between the confining length and
the deconfining temperature is what one would expect from strong
interaction and QCD physics; if the length of confinement of a
fermion is small, the energy required to overcome its confining
barrier is large, and vice-versa. Roughly speaking, one has a sort
of uncertainty relation $L \sim 1/p$, with $p$ being the momentum
which, for relativistic particles, is proportional to the energy.
Anyhow, the questions of how the fermions are confined and how they
get unconfined are not answered with our analysis.

Since we have completely determined the relevant dependence of
$L_{c}^{(D)}$ and $T_{d}^{(D)}$ on $\lambda$, for any value of
$\lambda$ ($\geq \lambda_{c}^{(D)}$), estimates of values of the
confining length and the deconfining temperature require the
specification of the mass of the fermions. Viewing the GN model as
an effective model for strong interaction, a natural choice is to
take the constituent quark mass, $m \approx 350 \, {\rm MeV} \simeq
1.75\,{\rm fm}^{-1}$ \cite{mass1}. With such a choice, we obtain the
limiting values of the confining length and the deconfining
temperature presented in Table~1.

\begin{table}[h]
\caption{Limiting values of $L_{c}^{(D)}(\lambda)$ $[{\rm fm}]$ and
$T_{d}^{(D)}(\lambda)$ $[{\rm MeV}]$, for $m \approx 350\,{\rm MeV}
\simeq 1.75\,{\rm fm}^{-1}$.}
\begin{tabular*}{8.6cm}{ccccc}
\hline \hline  $\,\,\,\,\,\,\,\,\,\,D\,\,\,\,\,\,\,\,\,$
\vspace{0.1cm} & $\,\,\,\,\,\,\,\,L_{{\rm
min}}^{(D)}\,\,\,\,\,\,\,\,$ & $\,\,\,\,\,\,\,\,L_{{\rm
max}}^{(D)}\,\,\,\,\,\,\,\,$ & $\,\,\,\,\,\,\,\,T_{{\rm
min}}^{(D)}\,\,\,\,\,\,\,\,$
& $\,\,\,\,\,\,\,\,T_{{\rm max}}^{(D)}\,\,\,\,\,\,\,\,$ \\
\hline
2 \vspace{0.1cm}& $0.45$ & $0.96$ & $227$ & $451$ \\
3 \vspace{0.1cm}& $0.74$ & $1.2$ & $189$ & $304$  \\
4 \vspace{0.1cm}& $0.96$ & $1.14$\footnote{These values of $L_{{\rm
max}}^{(4)}$ and $T_{{\rm min}}^{(4)}$ correspond to the redefined
value of $\lambda_{c}^{(4)}$.} & $193$$^{a}$ & $245$ \\
\hline \hline
\end{tabular*}
\end{table}

We find that the range of variation of $L_{c}^{(D)}$ and
$T_{d}^{(D)}$, as $\lambda$ runs in the strong coupling regime
($\geq \lambda_{c}^{(D)}$), is relatively small and decreases as $D$
increases. These values compare amazingly well with the size of
hadrons (e.g., the experimentally measured proton charge diameter is
$\approx 1.74\,{\rm fm}$~\cite{kars}) and the estimated deconfining
temperature ($\approx 200\,{\rm MeV}$) for hadronic
matter~\cite{temp1}.

To summarize, we have established the existence of a phase
transition in the massive large-$N$ GN model with compactified
spatial dimensions. It remains to be investigated other aspects of
this transition like effects of finite density and critical
exponents. Such a study is left for a future work.\\

\begin{center}
{\bf ACKNOWLEDGEMENTS}
\end{center}

This work was partially supported by CNPq and CAPES (Brazil), and
NSERC (Canada).

\end{document}